\documentclass[sigconf]{acmart}

\usepackage{caption}
\usepackage{balance}
\AtBeginDocument{%
  }

\setcopyright{acmlicensed}

\copyrightyear{2025} 
\acmYear{2025} 
\setcopyright{cc}
\setcctype{by}
\acmConference[SIGIR '25]{Proceedings of the 48th International ACM SIGIR Conference on Research and Development in Information Retrieval}{July 13--18, 2025}{Padua, Italy}
\acmBooktitle{Proceedings of the 48th International ACM SIGIR Conference on Research and Development in Information Retrieval (SIGIR '25), July 13--18, 2025, Padua, Italy}
\acmDOI{10.1145/3726302.3731966}
\acmISBN{979-8-4007-1592-1/2025/07}

\begin{document}

\title{Progressive Refinement of E-commerce Search Ranking Based on Short-Term Activities of the Buyer}

\author{Taoran Sheng}
\orcid{0000-0002-2760-2699}
\email{tsheng@ebay.com}
\affiliation{%
  \institution{eBay Inc.}
  \city{San Jose}
  \state{California}
  \country{USA}
}

\author{Sathappan Muthiah}
\orcid{0009-0009-3687-3392}
\email{smuthiah@ebay.com}
\affiliation{%
  \institution{eBay Inc.}
  \city{San Jose}
  \state{California}
  \country{USA}
  }

\author{Atiq Islam}
\orcid{0009-0002-1462-7392}
\email{atislam@ebay.com}
\affiliation{%
  \institution{eBay Inc.}
  \city{New York}
  \state{New York}
  \country{USA}
}

\author{Jinming Feng}
\orcid{0009-0000-8490-360X}
 \email{jinmfeng@ebay.com}
\affiliation{%
  \institution{eBay Inc.}
  \city{Shanghai}
  \country{China}
}

\renewcommand{\shortauthors}{Taoran Sheng, Sathappan Muthiah, Atiq Islam, and Jinming Feng}

\begin{abstract}
In e-commerce shopping, aligning search results with a buyer’s immediate needs and preferences presents a significant challenge, particularly in adapting search results throughout the buyer's shopping journey as they move from the initial stages of browsing to making a purchase decision or shift from one intent to another. This study presents a systematic approach to adapting e-commerce search results based on the current context. We start with basic methods and incrementally incorporate more contextual information and state-of-the-art techniques to improve the search outcomes. By applying this evolving contextual framework to items displayed on the search engine results page (SERP), we progressively align search outcomes more closely with the buyer’s interests and current search intentions. Our findings demonstrate that this incremental enhancement, from simple heuristic autoregressive features to advanced sequence models, significantly improves ranker performance. The integration of contextual techniques enhances the performance of our production ranker, leading to improved search results in both offline and online A/B testing in terms of Mean Reciprocal Rank (MRR). Overall, the paper details iterative methodologies and their substantial contributions to search result contextualization on e-commerce platforms.
\end{abstract}





\maketitle
\section{Introduction}
E-commerce search result pages are designed to seamlessly connect buyers with items that match their search intent, highlighting the critical role of search ranking algorithms in enhancing user experience. With diverse buyer cohorts and a vast inventory, achieving an optimal item ranking for search queries is challenging. Buyers have unique preferences that a universally trained ranking model often fails to address. To provide a pleasant shopping experience, search engines must learn and exploit individual preferences when presenting search results \cite{10.1145/3366424.3382715}. To address these complexities, various personalization techniques have been developed \cite{10.1145/3626772.3661366, 10.1145/3626772.3661364, 10.1145/3539618.3591842}, including content-based strategies \cite{cinar2020adaptive, lops2019trends, 10.1145/2959100.2959166}, user behavior-based tactics \cite{supersonalization, 10.1145/3366424.3386197, 10.1145/3397271.3401440}, and hybrid methods \cite{hui2022personalized, kouki2020generating}. These approaches often rely on user profiles requiring substantial behavioral data to capture personal preferences. Traditional user modeling aggregates interactions over time to construct a comprehensive representation of preferences \cite{10.1145/3340531.3411877}. However, these methods generally need significant interaction history, which can be restrictive. Moreover, the same buyer may purchase different types of products, and different buyers might use the same account. Even within the same search session, buyers may switch their intent and shop for seemingly different items. These factors highlight the limitations of relying solely on personalization and underscore the importance of understanding the buyer's current context and preferences. Contextualization offers a solution by utilizing short-term user behaviors, providing an alternative that doesn't depend on extensive historical data.

In this work, we progressively refine search contextualization techniques based on buyers' short-term activities, transitioning from simple to advanced methods. We integrate basic and complex features as contextual inputs in eBay's search ranker, starting with heuristic approaches and extending to intent-aware methods \cite{10.1145/3397271.3401440}. To enhance ranking, we incorporate complex features using sequential attention, inspired by their success in search ranking tasks \cite{bi2020transformer, 10.1145/3437963.3441667, laskar2020contextualized}. Our enhanced ranking model demonstrates the effectiveness of contextual signals in boosting MRR \cite{Craswell2009MeanRR}. We explore integrating these features into the ranking model without relying on long-term user profiles, discussing improvements in search ranking performance. Our contributions are as follows: we combine and compare various contextual features, from simple to advanced transformer-based features, illustrating how they align search ranking outcomes with buyers' desires, dramatically enhancing e-commerce search performance.

\section{Methodology}
This section outlines the methodologies used to improve search ranking contextualization. We begin with heuristic autoregressive features, leveraging the recent 1 to 5 clicks within the current session to provide immediate context. Next, we utilize the buyer's current query to make the contextualization intent-aware, thereby enhancing its effectiveness and improving search result relevance. Finally, we explore sequential attention-refined contextual features, capturing comprehensive context from the buyer's short-term click history. These features are designed for direct integration into the ranking model to boost contextualization and optimize search outcomes.

\subsection{Heuristic Autoregressive Contextualization} 
\label{basicfeat}
We developed ranking features that use buyer click interaction data over various time frames to contextualize search results. Heuristic autoregressive techniques model dependencies between past and current clicks, improving contextualization. These features link items on the current SERP with previous clicks, focusing on immediate and slightly extended time frames. We employ both textual and embedding-based methods to capture user preferences.

\subsubsection{Last Click}
To provide immediate contextual relevance, we use the user's most recent click.

Textual Feature: We apply normalized compression distance (NCD) \cite{cilibrasi2004clusteringcompression} to measure textual distance between current SERP item titles and the last clicked item's title. Here, \( n \) denotes the current time stamp, and \( n-1 \) denotes the most recent time stamp. For the \(i\)-th current SERP item \( \mathrm{item}_i^{n} \) and previous click \( \mathrm{item}^{n-1} \), with \( T \) as the title and \( C(T) \) as the compressed title size, the NCD distance is:
\[
\mathrm{NCD}_{i, \text{lc}} = \frac{C(T_{\mathrm{item}_i^{n}} + T_{\mathrm{item}^{n-1}}) - \min(C(T_{\mathrm{item}_i^{n}}), C(T_{\mathrm{item}^{n-1}}))}{\max(C(T_{\mathrm{item}_i^{n}}), C(T_{\mathrm{item}^{n-1}}))}
\]

Embedding Feature: We calculate cosine similarity between embeddings of current SERP items and the last clicked item, using eBert model embeddings \cite{DBLP:journals/corr/abs-2108-10197}. Here, \(\mathbf{E}_{\mathrm{item}_i^{n}}\) and \(\mathbf{E}_{\mathrm{item}^{n-1}}\) represent the embeddings of the \( i \)-th current SERP item and the last clicked item, respectively. The cosine similarity is calculated as:
\[
\mathrm{CosSim}_{i, \text{lc}} = \frac{\mathbf{E}_{\mathrm{item}_i^{n}} \cdot \mathbf{E}_{\mathrm{item}^{n-1}}}{\|\mathbf{E}_{\mathrm{item}_i^{n}}\| \|\mathbf{E}_{\mathrm{item}^{n-1}}\|}
\]

\(\mathrm{NCD}_{i, \text{lc}}\) and \(\mathrm{CosSim}_{i, \text{lc}}\) indicate the alignment between the current result \( \mathrm{item}_i^{n} \) and the user's immediate past interests.

\subsubsection{Last 5 Clicks}
To capture broader user preferences, we use data from the user's last five clicks.

Textual Feature: We extend the \(\mathrm{NCD}\) by concatenating the titles of the last five clicks into a single text string, \( T_{\mathrm{concat}} \), providing a comprehensive view of recent interests. The \(\mathrm{NCD}\) for current SERP items is calculated as:
\[
\mathrm{NCD}_{i, \text{l5c}} = \frac{C(T_{\mathrm{item}_i^{n}} + T_{\mathrm{concat}}) - \min(C(T_{\mathrm{item}_i^{n}}), C(T_{\mathrm{concat}}))}{\max(C(T_{\mathrm{item}_i^{n}}), C(T_{\mathrm{concat}}))}
\]

Embedding Feature: We calculate the cosine similarity between embeddings of current SERP items and each of the last five clicked items, averaging these values for overall relevance. Here, \( \mathrm{item}^{n-j} \) denotes the \( j \)-th most recent click, where \( j \) ranges from 1 to 5:
\[
\mathrm{CosSim}_{i, \text{l5c}} = \frac{1}{5} \sum_{j=1}^{5} \frac{\mathbf{E}_{\mathrm{item}_i^{n}} \cdot \mathbf{E}_{\mathrm{item}^{n-j}}}{\|\mathbf{E}_{\mathrm{item}_i^{n}}\| \|\mathbf{E}_{\mathrm{item}^{n-j}}\|}
\]

By structuring features around different time frames, we compare and balance immediate relevance with broader user preferences.

% \subsection{Retrieval-Based Personalization} \label{retrievefeat}
\subsection{Intent-Aware Contextualization} \label{retrievefeat}
\begin{figure}[!ht]
    \vspace{-7pt} 
    \centering
    \begin{minipage}{0.41\textwidth}
        \includegraphics[width=\linewidth]{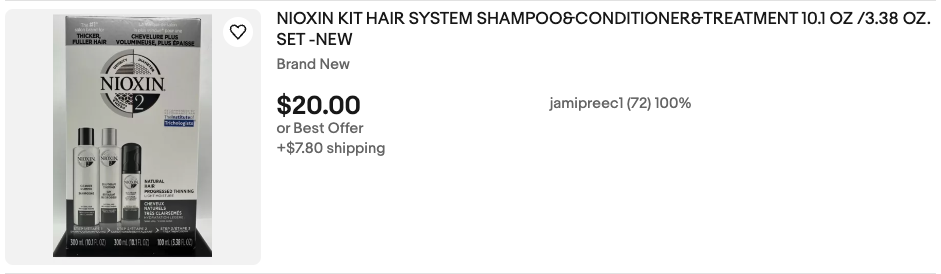}
    \end{minipage}
    \begin{minipage}{0.41\textwidth}
        \includegraphics[width=\linewidth]{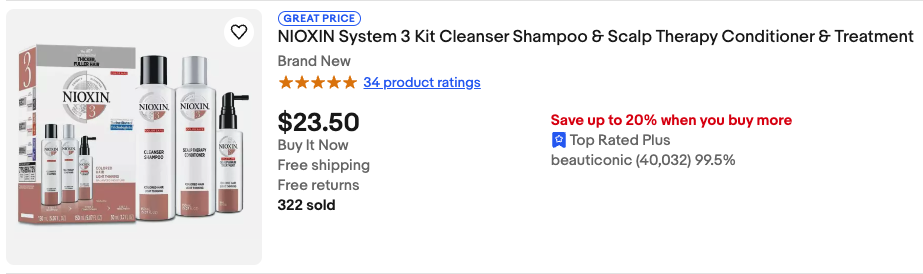}
    \end{minipage}
    \begin{center}
        \small (a) SERP with heuristic autoregressive contextualization.
    \end{center}
    \begin{minipage}{0.41\textwidth}
        \includegraphics[width=\linewidth]{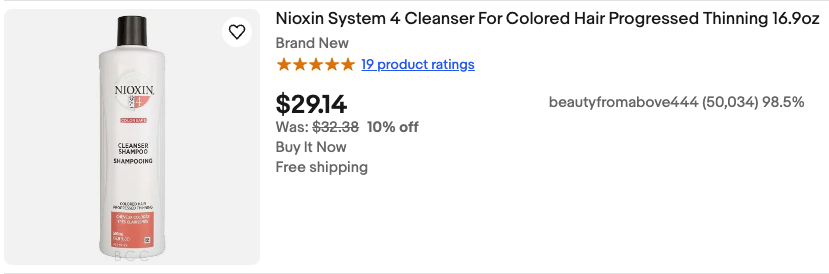}
    \end{minipage}
    \begin{minipage}{0.41\textwidth}
        \includegraphics[width=\linewidth]{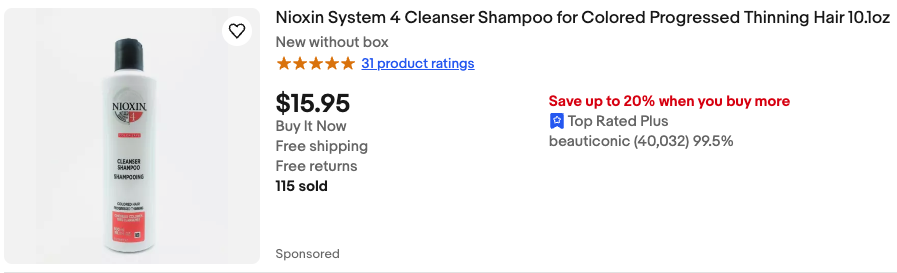}
    \end{minipage}
    \begin{center}
        \small (b) SERP with intent-aware contextualization.
    \end{center}
    \vspace{-7pt} 
    \caption{Different contextualized search results for query \textit{"Nioxin system 4 colored hair progressed thinning kit"}.}
    \label{introExample}
\end{figure}

Heuristic autoregressive features use recent click history to infer buyer preferences, but they may not align with the user's current search intent. For example, when a user queries \textit{"Nioxin system 4 colored hair progressed thinning kit"}, relying solely on recent clicks can mislead search results, ranking less relevant items higher, as shown in Figure \ref{introExample}(a).

To address this, we propose an intent-aware contextualization strategy that combines past clicks with the current query to dynamically refine context. As shown in Figure \ref{introExample}(b), this approach enhances search results by better aligning them with the user's ongoing intent. This strategy improves the effectiveness of ranking models in e-commerce by accurately capturing the user's current preferences. Our method selectively focuses on clicks most pertinent to the current query, avoiding irrelevant context. This targeted approach enhances the precision of contextualization. We developed two new features using textual and embedding-based measures to better capture user preferences.

\subsubsection{Intent-Aware Textual Feature}
This feature enhances contextualization by using textual distance to compare the user's current query with titles from the last five historical clicks. The item with the smallest NCD distance is chosen as the reference for the buyer's current intent. We then apply NCD again to compare this reference item's title with current SERP item titles, focusing on direct textual relevance to improve result relevance. The NCD for selecting the reference item based on the query is:
\[
\mathrm{NCD}_{\text{query, item}^{n-j}} = \frac{C(Q + T_{\mathrm{item}^{n-j}}) - \min(C(Q), C(T_{\mathrm{item}^{n-j}}))}{\max(C(Q), C(T_{\mathrm{item}^{n-j}}))}
\]

where \( Q \) is the current query and \( T_{\mathrm{item}^{n-j}} \) is the title of the \( j \)-th most recent historical click item. The reference item is chosen as:
\[
\text{Ref}_{\text{txt}} = \arg\min_{j} \left( \mathrm{NCD}_{\text{query, item}^{n-j}} \right)
\]

% Once the reference item is chosen, the NCD between its title and current SERP item titles is:
Once the reference item is chosen, the NCD between its title \( T_{\mathrm{Ref}_{\text{txt}}} \) and current SERP item titles is calculated as follows:
\[
\mathrm{NCD}_{i, \text{Ref}_{\text{txt}}} = \frac{C(T_{\mathrm{item}_i^{n}} + T_{\mathrm{Ref}_{\text{txt}}}) - \min(C(T_{\mathrm{item}_i^{n}}), C(T_{\mathrm{Ref}_{\text{txt}}}))}{\max(C(T_{\mathrm{item}_i^{n}}), C(T_{\mathrm{Ref}_{\text{txt}}}))}
\]
% Here, \( T_{\mathrm{Ref}_{\text{txt}}} \) is the reference item's title.

\subsubsection{Intent-Aware Embedding Feature}
This feature employs a semantic approach to contextualization using embeddings from the eBert model. We compute cosine similarity between the current query embedding and embeddings of items from the user's last five clicks. The item with the highest similarity serves as the reference for the buyer's current intent. The cosine similarity for selecting the reference item based on the query is calculated as:
\[
\text{CosSim}_{\text{query, item}^{n-j}} = \frac{\mathbf{E}_Q \cdot \mathbf{E}_{\mathrm{item}^{n-j}}}{\|\mathbf{E}_Q\| \|\mathbf{E}_{\mathrm{item}^{n-j}}\|}
\]

where \(\mathbf{E}_Q\) is the query embedding, and \(\mathbf{E}_{\mathrm{item}^{n-j}}\) is the embedding of the \( j \)-th most recent historical click item. The reference item is chosen as:
\[
\text{Ref}_{\text{emb}} = \arg\max_{j} \left( \text{CosSim}_{\text{query, item}^{n-j}} \right)
\]

Once the reference item is determined, we calculate the cosine similarity between this reference item's embedding and the current SERP item embeddings:
\[
\text{CosSim}_{i, \text{Ref}_{\text{emb}}} = \frac{\mathbf{E}_{\mathrm{item}_i^{n}} \cdot \mathbf{E}_{\mathrm{Ref}_{\text{emb}}}}{\|\mathbf{E}_{\mathrm{item}_i^{n}}\| \|\mathbf{E}_{\mathrm{Ref}_{\text{emb}}}\|}
\]

Here, \(\mathbf{E}_{\mathrm{Ref}_{\text{emb}}}\) is the reference item's embedding, and \(\mathbf{E}_{\mathrm{item}_i^{n}}\) is the embedding of the \( i \)-th SERP item. This method captures semantic relationships and user preferences beyond text, enriching search result contextualization.

\subsection{Sequential Attention Contextualization} \label{seqfeat}
The intent-aware contextual features discussed in Section \ref{retrievefeat} focus on a single relevant past interaction, potentially missing the broader relevance of multiple interactions throughout the buyer's shopping journey. To address this, we introduce two sequence models with attention mechanisms designed to enhance contextualization by effectively capturing the user's short-term click history. These models process sequences of clicks, selectively emphasizing those clicks that are more relevant to the evolving context of the purchase process, while diminishing the influence of less pertinent clicks.

The sequence models are trained to generate embeddings that encapsulate nuanced patterns in user behavior, reflecting the dynamic nature of the shopping experience. We implement these features by computing cosine similarities between the historical embeddings generated by the sequence models and the current SERP item embeddings:
\[
\text{CosSim}_{i, \text{seq}} = \frac{\mathbf{E}_{\mathrm{item}_i^{n}} \cdot \mathbf{E}_{\text{seq}}}{\|\mathbf{E}_{\mathrm{item}_i^{n}}\| \|\mathbf{E}_{\text{seq}}\|}
\]

where \(\mathbf{E}_{\text{seq}}\) is the historical embedding generated from the sequence model based on the last 5 historical clicks.
%, and \(\mathbf{E}_{\mathrm{item}_i^{n}}\) is the embedding of the \(i\)-th SERP item. 
Following this, we will discuss how the historical embedding \(\mathbf{E}_{\text{seq}}\) is generated using different sequence models.

\subsubsection{Transformer-Generated Embedding}
The first model uses a standard transformer architecture to process the user's click history. The transformer's attention mechanisms allow it to weigh different parts of the input sequence based on their importance within the user's historical interactions. By processing a sequence of eBert click embeddings, \(\{\mathbf{E}_{item^{n-5}}, \mathbf{E}_{item^{n-4}}, \ldots, \mathbf{E}_{item^{n-1}}\}\), which only includes past clicks and excludes the current query, the model produces a unified embedding \(\mathbf{E}_{\text{seq: trans}}\) that captures sequential dependencies and contextual nuances. This historical embedding is generated as follows:
\[
\mathbf{E}_{\text{seq: trans}} = \text{Transformer}(\{\mathbf{E}_{item^{n-5}}, \mathbf{E}_{item^{n-4}}, \ldots, \mathbf{E}_{item^{n-1}}\})
\]

To align with search ranking objectives, the transformer is trained with a lambdaRank loss function \cite{burges2010ranknet}, prioritizing sale items over clicks and clicks over items with no engagement. The training data is collected from different time frames than the ranking model's data to avoid data leakage. This comprehensive user preference representation aids in contextualizing search results by aligning rankings with historical interests and behaviors.

\subsubsection{Perceiver-Generated Embedding}
The second model is a lightweight adaptation of the Perceiver architecture \cite{DBLP:journals/corr/abs-2103-03206, DBLP:journals/corr/abs-2107-14795}, designed to efficiently integrate query information with the user's click history in two stages. Initially, the model computes cross-attention between the eBert query embedding \(\mathbf{Q}\) and the sequence of eBert click embeddings \(\{\mathbf{E}_{item^{n-5}}, \mathbf{E}_{item^{n-4}}, \ldots, \mathbf{E}_{item^{n-1}}\}\). Cosine similarity is used for its efficiency and ability to capture relationships between the query and user history. This adaptation simplifies the traditional Perceiver model by omitting the latent array, reducing computational overhead for real-time applications. The cross-attention output \(\mathbf{E_{\text{seq: cross}}}\) is computed as:
\[
\mathbf{E_{\text{seq: cross}}} = \text{CrossAttention}(\mathbf{Q}, \{\mathbf{E}_{item^{n-5}}, \mathbf{E}_{item^{n-4}}, \ldots, \mathbf{E}_{item^{n-1}}\})
\]

Next, self-attention is applied to \(\mathbf{E_{\text{seq: cross}}}\), refining the integration of query and click history, similar to the Perceiver's iterative refinement:
\[
\mathbf{E}_{\text{seq: perc}} = \text{SelfAttention}(\mathbf{E_{\text{seq: cross}}})
\]

This model is also trained with a lambdaRank loss function, using data from different time periods to prevent data leakage, ensuring the training data is separate from the data used by the ranking model. The Perceiver-like model focuses on relevant interactions and filters out noise, generating an embedding that aligns with the user's current search intent, thereby enhancing the contextualization of search results.

\section{Evaluation}
\subsection{Ranking Model}
In our experiments, we use the eBay machine-learned ranking model (MLR) as the baseline. All rankers are trained on data from two weeks of eBay search logs, where each session includes a sequence of clicked items, serving as the basis for contextual features. For a query \( q \) from a buyer \( b \), the search engine returns a set of items \( SER_{q, b} \). Each \( item_i \) in the \(SER_{q, b}\) is characterized by features \( f_i \) and a score \( s_i \), reflecting engagement levels: sale items score higher than clicked items, and clicked items score higher than unengaged items. The goal is to train a ranker \( R \) to predict the score \( s_i \) for each \( item_i \) based on its features \( f_i \). This process involves minimizing a lambdaRank loss function \(\sum_{item_i \in SER_{q, b}} Loss(R(f_i), s_i)\). Here, \( R(f_i) \) is the baseline ranker, while the contextualized ranker is \( R(f_i + f^{\text{context}}_i) \), where \( f^{\text{context}}_i \) are the contextual features.

We incorporate features from Sections \ref{basicfeat}, \ref{retrievefeat}, and \ref{seqfeat} into the MLR for contextualization. The process involves three main steps, each exploring multiple variants. First, we integrate heuristic autoregressive contextual features into the MLR and compare each variant to the original MLR. Next, we develop variant \textbf{P1} by adding these autoregressive features to the MLR, then refine \textbf{P1} with intent-aware contextual features, comparing each variant to \textbf{P1}. Finally, we enhance \textbf{P1} with intent-aware contextual features to create \textbf{P2}, and introduce sequential attention contextual features into \textbf{P2} to capture comprehensive user preferences, comparing results to \textbf{P2}. This systematic approach progressively improves the MLR's ability to deliver contextualized search experiences, demonstrating that each addition and variant exploration improves performance.

\subsection{Offline Evaluation}
For offline evaluation, we analyze the impact of various contextualized MLR variants on the MRR of sale items. Table \ref{offline_mlr_lift_comparison} summarizes the MRR Sale improvements achieved by different contextualized MLR variants compared to their respective baselines.
\begin{table}[!ht]
    \vspace{-7pt}
    \centering
    \caption{Offline MRR Sale Lift Comparison for Different Variants}
    \vspace{-7pt}
    \resizebox{0.38\textwidth}{!}{%
    \begin{tabular}{|l|c|}
        \hline
        \textbf{Variant} & \textbf{MRR Sale Lift} \\
        \hline
        \multicolumn{2}{|l|}{\textbf{Heuristic Autoregressive Contextual Features}} \\
        \hline
        MLR + $\mathrm{NCD}_{\text{lc}}$ vs MLR & 1.16\% \\ 
        MLR + $\mathrm{CosSim}_{\text{lc}}$ vs MLR & 1.84\% \\
        MLR + $\mathrm{NCD}_{\text{l5c}}$ vs MLR & 0.60\% \\ 
        MLR + $\mathrm{CosSim}_{\text{l5c}}$ vs MLR & 1.23\% \\ 
        \hline
        \multicolumn{2}{|l|}{\textbf{Intent-Aware Contextual Features}} \\
        \hline
        P1 + $\mathrm{NCD}_{\text{Ref}_{\text{txt}}}$ vs P1 & 0.93\% \\ 
        P1 + $\text{CosSim}_{\text{Ref}_{\text{emb}}}$ vs P1 & 1.08\% \\
        \hline
        \multicolumn{2}{|l|}{\textbf{Sequential Attention Contextual Features}} \\
        \hline
        P2 + $\text{CosSim}_{\text{seq: trans}}$ vs P2 & 0.69\% \\ 
        P2 + $\text{CosSim}_{\text{seq: perc}}$ vs P2 & 1.01\% \\ 
        \hline
    \end{tabular}
    }
    \label{offline_mlr_lift_comparison}
    \vspace{-7pt}
\end{table}

The results in Table \ref{offline_mlr_lift_comparison} reveal several insights. Embedding similarity with the previous click yields a 1.84\% improvement, while textual NCD with the last 5 clicks shows a 0.60\% lift, suggesting that embedding-based features more effectively capture user preferences. Extended click histories may introduce irrelevant items, misaligning with the user's current intent. Both textual and embedding-based intent-aware contextual features show similar performance enhancements over P1, indicating that retrieving relevant items from the user's click history can reduce noise from irrelevant clicks and clarify the buyer's current search intent. The Perceiver-based feature outperforms the transformer-based feature, with a 1.01\% lift over P2, suggesting that the Perceiver-like model's iterative attention mechanism effectively integrates query and click history for refined contextualization. Overall, these findings demonstrate that progressively incorporating sophisticated contextual features leads to measurable improvements in search ranking performance, highlighting the potential of selective approaches in capturing user preferences and enhancing search result contextualization on e-commerce platforms.

\subsection{Online Evaluation}
For online experiments, we conducted A/B testing over two weeks to evaluate the contextualized MLR variants across all platforms, including desktop, mobile web, iOS, and Android. Users were randomly assigned to either the control or treatment group, with equal distribution of traffic. Table \ref{online_mlr_lift_comparison} presents the results for the three variants with completed experiments, while testing continues for the remaining variants.
\begin{table}[!ht]
    \vspace{-7pt}
    \centering
    \caption{Online MRR Sale Lift Comparison for Different Variants}
    \vspace{-7pt}
    \resizebox{0.38\textwidth}{!}{%
    \begin{tabular}{|l|c|}
        \hline
        \textbf{Variant} & \textbf{MRR Sale Lift} \\
        \hline
        \multicolumn{2}{|l|}{\textbf{Heuristic Autoregressive Contextual Features}} \\
        \hline
        MLR + $\mathrm{NCD}_{\text{lc}}$ vs MLR & 1.30\% \\ 
        MLR + $\mathrm{NCD}_{\text{l5c}}$ vs MLR & 0.72\% \\ 
        \hline
        \multicolumn{2}{|l|}{\textbf{Intent-Aware Contextual Features}} \\
        \hline
        P1 + $\mathrm{NCD}_{\text{Ref}_{\text{txt}}}$ vs P1 & 0.96\% \\ 
        % Embedding Similarity Retrieved Click vs P1 & 1.08\% \\
        \hline
    \end{tabular}
    }
    \label{online_mlr_lift_comparison}
    \vspace{-7pt}
\end{table}

The online evaluation results in Table \ref{online_mlr_lift_comparison} demonstrate the effectiveness of contextualized MLR variants in real-world settings. The autoregressive contextual feature can improve MRR Sale by 1.30\%, showing that even simple contextualization can significantly enhance user experience by aligning search results with recent interactions. Additionally, an intent-aware contextual feature can achieve a 0.96\% improvement further, indicating that synchronizing buyers' past interactions with their current intent enhances search ranking outcomes. These results confirm that contextual features improve search performance and user satisfaction on e-commerce platforms. Ongoing testing, particularly with advanced model features, is expected to provide deeper insights, further refining the contextualization strategy across various search scenarios.

\section{Conclusion and Future Work}
This work systematically compared and analyzed various contextual features to refine search ranking on e-commerce platforms, ranging from heuristic autoregressive features to intent-aware features and sequence model-based features. Both offline and online evaluations demonstrated significant improvements in MRR for sale items, underscoring the effectiveness of these approaches in capturing user preferences. The findings highlight the potential of advanced contextualization techniques to improve user satisfaction by delivering more tailored search results. Future work could focus on leveraging diverse interaction types beyond buyer clicks for richer contextualization.

\bibliographystyle{ACM-Reference-Format}
\balance
\bibliography{sample-base}

% \appendix









\end{document}